\documentclass[a4paper,fleqn,usenatbib]{mnras}

\usepackage{newtxtext,newtxmath}
\pdfoutput=1
\usepackage[T1]{fontenc}
\usepackage{ae,aecompl}

\usepackage{graphicx}	
\usepackage{amsmath}	
\usepackage{amssymb}	

\title[HST/STIS spectroscopy of F14394+5332]{Quantifying the AGN-driven outflows in ULIRGs (QUADROS) IV: HST/STIS 
spectroscopy of the sub-kpc warm outflow in F14394+5332}

\author[C. N. Tadhunter et al.]{
C. Tadhunter,$^{1}$\thanks{E-mail: c.tadhunter@sheffield.ac.uk},
L. Holden,$^{1}$, C. Ramos Almeida$^{2,3}$, D. Batcheldor$^4$
\\
$^{1}$ Department of Physics \& Astronomy, University of Sheffield, Sheffield S6 3TG, UK\\
$^2$Instituto de Astrof\' isica de Canarias, Calle V\' ia L\'actea, s/n, E-38205, La Laguna, Tenerife, Spain \\
$^3$Departamento de Astrof\' isica, Universidad de La Laguna, E-38205, La Laguna, Tenerife, Spain \\
$^4$ Physics and Space Sciences Department, Florida Institute of Technology, 150 West University Boulevard, Melbourne, FL 32901, USA \\
}

\date{Accepted XXX. Received YYY; in original form ZZZ}

\pubyear{2018}

\begin{document}
\label{firstpage}
\pagerange{\pageref{firstpage}--\pageref{lastpage}}
\maketitle

\begin{abstract}
Considerable uncertainties remain about the nature of warm, AGN-driven outflows and their impact on the evolution of galaxies. This is because the outflows are often unresolved in ground-based observations. As part of a project to study the AGN outflows in some of the most rapidly evolving galaxies in the local Universe, here we present HST/STIS observations of F14394+5332E that resolve the sub-kpc warm outflow for the first time in a ULIRG. The observations reveal a compact, high ionization outflow region ($r_{max}\sim$0.9\,kpc) set in a more extensive ($r_{max}\sim$1.4\,kpc) halo that is kinematically quiescent and has a lower ionization state. A large line width ($600 < FWHM < 1500$ km s$^{-1}$) is measured throughout the outflow region, and the outflowing gas shows a steep velocity gradient with radius, with the magnitude of the blueshifted velocities increasing from $\sim$500 to 1800\,km s$^{-1}$ from the inner to the outer part of the outflow. We interpret the observations in terms of the local acceleration, and hydrodynamic destruction, of dense clouds as they are swept up in a hot, low density wind driven by the AGN. We discuss the implications for measuring the mass outflow rates and kinetic powers for the AGN-driven outflows in such objects.
\end{abstract}




\section{Introduction}

Warm, AGN-driven outflows are potentially important for influencing the star formation histories 
of galaxies as they evolve \citep[e.g.][]{veilleux05,king15}. Therefore it is crucial that their properties are accurately quantified, so that their
true destructive potential can be gauged. 

\begin{figure*}
\includegraphics[width=15.0cm]{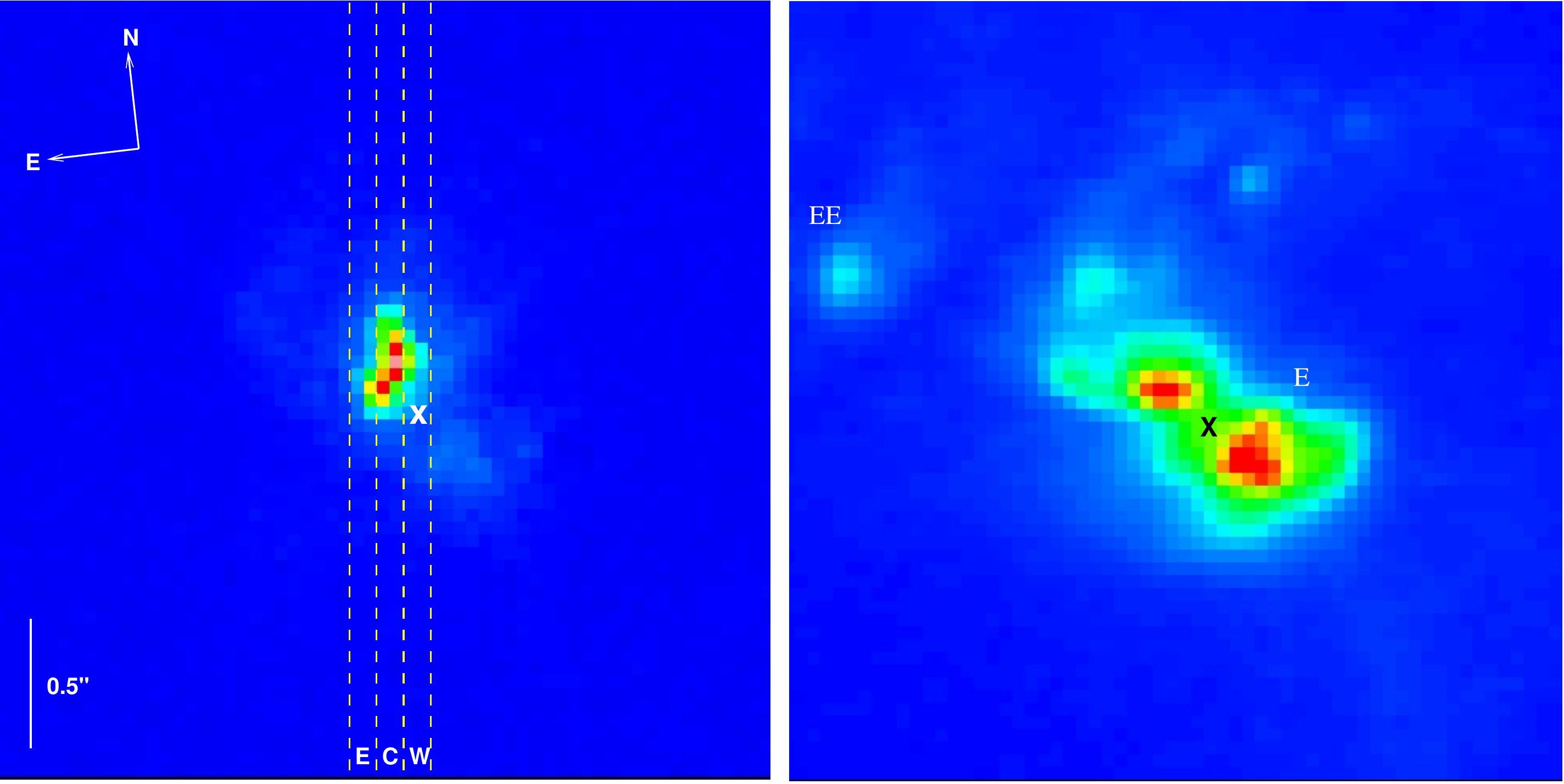}
\caption{Narrow-band [OIII]$\lambda\lambda$5007,4959 (left) and intermediate-band continuum (right) ACS/WFC HST images of F14394+5332. The [OIII] image was taken with the FR551N ramp filter ($\Delta \lambda = 82$\AA) centred on 5510\AA\,, while the continuum image was taken with the F647M ramp filter ($\Delta \lambda = 367$\AA) centred on 5967\AA\,  \citep[see][for details]{tadhunter18}. The three slit positions used for the STIS observations are indicated in the left-hand image, and the crosses indicate the assumed position of the AGN in F14394+5332E. The approximate positions of the E and EE components from \citet{kim02} are marked in the continuum image. Both images have the same scale and orientation.}
\label{fig:slits}
\end{figure*}

Using optical spectroscopic data, it has proved challenging in  the past to
determine key properties of the warm outflows such as their mass outflow rates and kinetic powers \citep[see discussion in][]{rz13,harrison18}. On the one hand, measuring the electron density of the outflowing gas -- key for determining accurate outflow gas masses -- has suffered from the fact that the traditional [OII](3726/3729) and [SII](6717/6731) diagnostic ratios are hard to measure, due to blending issues in the face of complex emission line profiles. On the other hand, due to the fact that the warm outflows are often unresolved, or poorly resolved, in ground-based observations for all but the lowest redshift AGN, considerable uncertainties remain about their radial extents, geometries, and the volumes of the host galaxies that they directly affect. 

To overcome these difficulties, we are undertaking a project that is aimed at directly quantifying the warm outflows in a sample of  nearby Ultra-luminous Infrared Galaxies \citep[ULIRGs][]{sanders96} with optically-detected AGN: the Quantifying Ulirg Agn-DRiven OutflowS (QUADROS) project \citep[see][for a description of the full sample]{rose18}. ULIRGs are particularly important in the context of AGN outflows, because they represent the most rapidly evolving galaxies in the local universe: objects in which both the galaxy bulges and supermassive black holes are growing rapidly due to the deposition of large masses of gas into the circum-nuclear regions as a result of merger-induced gas flows. Hydrodynamic simulations of the major, gas-rich galaxy mergers predict that AGN-induced outflows should be particularly important in such objects
\citep[e.g.][]{dimatteo05,johansson09}. 

In past papers on the QUADROS project we used [OII] and [SII] trans-auroral emission line ratios -- an alternative to the traditional [OII](3726/3729) and [SII](6717/6731) ratios -- to demonstrate that the electron densities in the outflow regions are substantially higher than assumed in many past studies of the warm outflows in AGN \citep[][papers I and III respectively]{rose18,spence18}. We also used a combination of ground-based spectra and HST narrow-band [OIII] imaging to show that the outflow regions are relatively compact
\citep[$0.06 < r < 5$\,kpc:][paper II]{tadhunter18}. Putting this information together, we found that, even in these rapidly evolving systems, the mass outflow rates are relatively modest ($0.1 < \dot{M} < 20$\,M$_{\odot}$ yr$^{-1}$), and the kinetic powers represent only a small fraction (0.01 -- 3\%)
of the bolometric luminosities of the AGN. This is potentially at odds with some of the theoretical studies which require that as much as   $\sim$5 -- 10\% of the accretion power of the AGN be deposited in the outflows (e.g. \citealt{fabian99}; \citealt{dimatteo05}; \citealt{johansson09}; but see \citealt{harrison18}).

Some of the apparent discrepancy between theory and observation may be due to the fact that the warm gas -- as detected via optical and near-IR emission lines -- represents only a small fraction of the total mass of the AGN outflows, with larger masses and kinetic powers contained in cooler neutral and molecular gas phases \citep[e.g.][]{rupke13,cicone14,gonz17,morganti15,fiore17,veilleux17}. However, in terms of the warm outflows studied in the QUADROS project, some uncertainties remain about the outflow properties -- particularly their radial extents and geometries -- because of a lack of observations that {\it directly} resolve the emission line kinematics.  Therefore, in this paper we present high-spatial-resolution spectroscopic observations taken with the Space Telescope Imaging Spectrograph (STIS) on board the Hubble Space Telescope (HST) of the nearby ULIRG F14394+5332. These observations resolve
a sub-kpc warm outflow for the first time in a ULIRG. The observations are used to investigate the geometry and radial extent of the outflow, and examine the implications for the determination of the outflow properties in more typical cases in which the outflow regions are are much less well-resolved.

\begin{figure*}
\includegraphics[width=15.0cm]{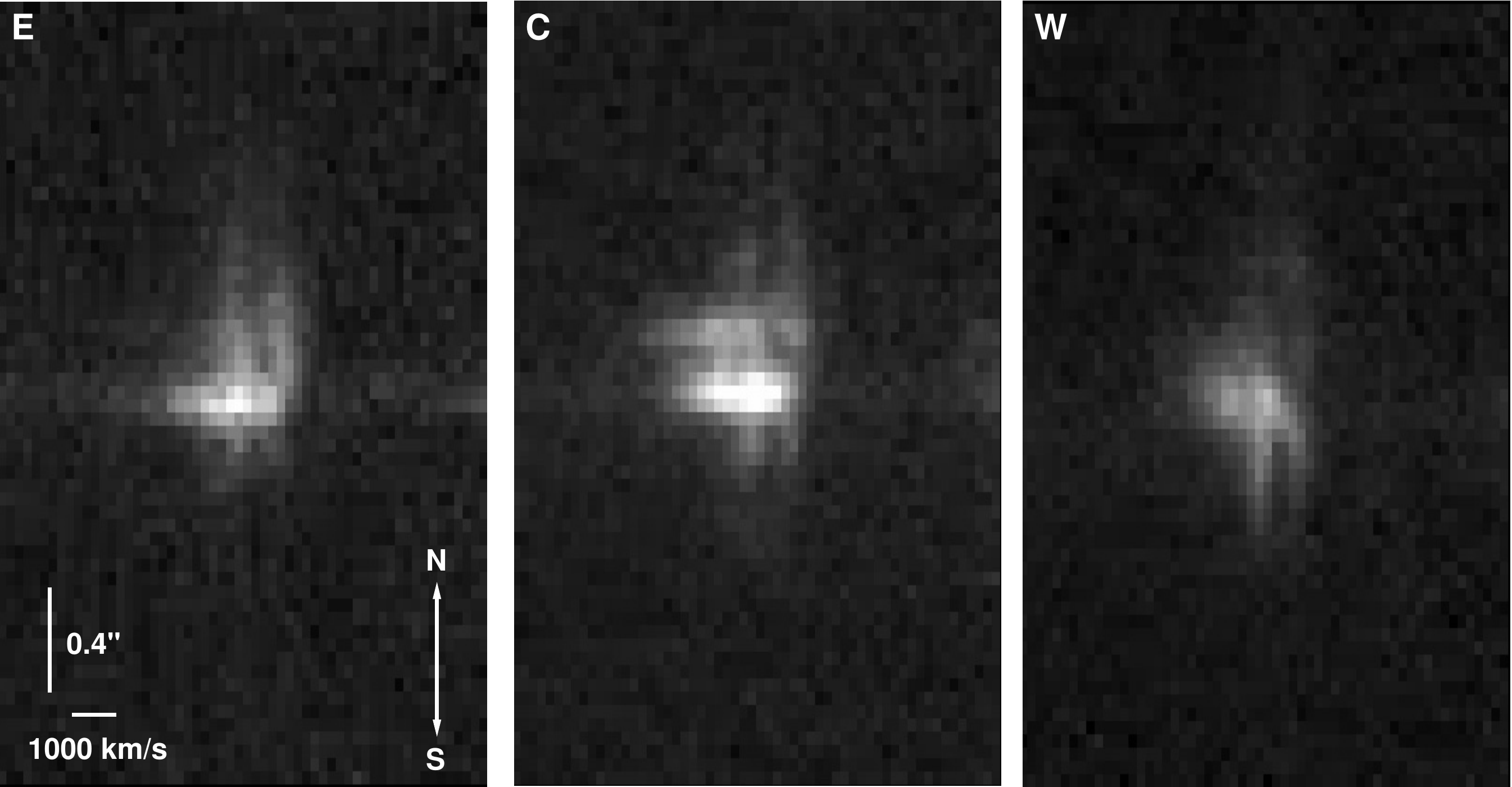}
\caption{Cut-outs of the two-dimensional, long-slit spectra of F14394+5332E for the three slit positions (E,C,W) showing the region of the H$\alpha$+[NII]$\lambda\lambda$6584,6548 blend. The x-axis represents wavelength (velocity) and the y-axis spatial position along the slit.}
\label{fig:2d}
\end{figure*}

F14394+5332 ($z = 0.10517$; \citealt{spence18}) is particularly interesting because it shows some of the most extreme emission line kinematics yet detected in a ULIRG \citep{lipari03,rz13,spence18}, with the [OIII] emission line profile dominated by multiple kinematic components that are  highly blueshifted ($\Delta V > 600$\,km s$^{-1}$) and/or broad ($FWHM > 1500$\,km s$^{-1}$). Based on its spatially integrated emission line properties, and estimates of the extent of the outflow determined from both the ground-based spectra and HST narrow-band [OIII] images, its warm outflow also shows one of the highest mass outflow rates ($1.4 < \dot{M} < 3.4$\,M$_{\odot}$ yr$^{-1}$) and kinetic powers as a fraction of AGN bolometric
luminosity ($0.12 < \dot{E}/L_{bol} < 0.66$\%) in the QUADROS sample. This object was selected for STIS observations because it is one of the 
few local ULIRGs for which our HST imaging observations indicate that the warm outflow region is both spatially resolved and has a sufficiently high surface brightness to allow spectroscopic observations with STIS.

Throughout this paper we assume a cosmology with H$_{0}$ = 73 km s$^{-1}$, $\Omega_{\rm m}$ = 0.27 and
$\Omega_{\Lambda}$ = 0.73, resulting in an angular-to-linear conversion factor of 1.853 kpc/arcsec 
at the redshift of F14394+5332.

\section{Observations and reductions}

HST/STIS observations were taken in Cycle 20 as part of program GO:12934 (PI Tadhunter), using the G750L grating to cover both the redshifted [OIII]$\lambda\lambda$5007,4959 and 
H$\alpha$+[NII]$\lambda\lambda$6584,6548 emission-line blends. With the 52$\times$0.1 slit aligned along PA\,172.1 we covered the bulk of the high-surface-brightness part of the extended [OIII] structure using three parallel
slit positions, each separated by 0.1 arcsec in the direction perpendicular to the slit (roughly the E-W direction).
We label these slits E (east), C (central), and W (west), and the positions of the slits are shown superimposed on our [OIII] narrow-band image \citep[see][for details]{tadhunter18} in Figure \ref{fig:slits}, where for comparison we also
show an intermediate-band, line-free continuum image. The total exposure times for the spectroscopic observations in the three slit positions were 2000, 2900 and 2900s for the E, C and W slits respectively (one orbit each including overheads). For each slit position
two exposures were taken at each of two along-the-slit dither positions separated by 1.5484 arcsec.

The preliminary reduction of the data was carried out using the standard CALSTIS pipeline, then residual hot pixels and
cosmic rays were removed using the CLEAN algorithm in the STARLINK FIGARO package, before the spectra from the two dither
positions were shifted and combined. The resulting pixel scales in the spectral and spatial direction were 4.882\AA/pixel, and 0.05078 arcsec/pixel respectively, with the latter significantly under-sampling the spatial line
spread function for the redshifted wavelength of [OIII] ($FWHM_{lsf} = 0.080\pm0.002$ arcsec; \citealt{spence18}). Given the diffuse nature of [OIII] outflow region detected in our narrow-band images, it is likely that the outflow fills the slit at any given location. In this case, according to the STIS manual, the instrumental width in the spectral direction is expected to be in the range $2 < FWHM < 3$ pixels ($9.7 < FWHM < 14.6$ \AA). Indeed, the weighted mean width of the H$\alpha$+[NII]$\lambda\lambda$6584,6548 lines measured in spatially extended regions to the south of the main outflow region along the
Western slit (this is where the lines appear most narrow) is 14.41$\pm$0.28 \AA. We adopt the latter as the instrumental width in subsequent calculations, although we note that this may slightly overestimate the true instrumental width if the intrinsic velocity width in the non-outflow regions is significant.

Following the extraction of 1D spectra for regions of interest along the slit, the emission line fluxes, widths and wavelength centroids were measured by fitting Gaussian profiles using the STARLINK DIPSO package. In the case of the [OIII]$\lambda\lambda$5007,4959 and [NII]$\lambda\lambda$6584,6548 doublets,  the doublet ratios and wavelength separations were set to the values required by atomic physics, and the widths of particular kinematic components were set equal for the two doublet components. In addition, in an attempt to avoid degeneracies in the fits to the H$\alpha$+[NII] blend given the relatively low spectral resolution, the widths and velocity shifts of individual kinematics components of H$\alpha$ were constrained to be the same as those of the [NII]$\lambda\lambda$6584,6548 lines. Whereas  single (generally broad) Gaussians provided adequate fits to the [OIII]$\lambda\lambda$5007,4959 profiles in all regions, two Gaussian components (broad+narrow) were required to fit the H$\alpha$ and [NII]$\lambda\lambda$6584,6548 lines in some parts of the outflow region. 

The F14394+5332 system has a complex structure on a large scale, which includes a spiral galaxy to the west that is connected by a tidal bridge to a more disturbed galaxy $\sim$52\,kpc to the east \citep{kim02,tadhunter18}. Based on near-IR imaging \citep{kim02}, this eastern galaxy has two nuclei -- labelled F14394+5332E and F14394+5332EE -- separated by $\sim$1.9 arcsec ($\sim$3.5\,kpc), but the [OIII] emission and AGN are associated with the most westerly of these two nuclei (F14394+5332E). Although the continuum structure of F14394+5332E is complex at optical wavelengths, there are two main continuum condensations bisected by a dust lane, which also bisects the [OIII] structure. Therefore, given the status of F14394+5332E as a type II AGN, it is plausible that its nucleus is hidden in the dust lane -- somewhere close to the position of the cross in Figure \ref{fig:slits}. However, lacking the high resolution mid-IR imaging that might locate the nucleus more accurately, it is clear that there remains some uncertainty about the true nuclear position. In what follows we will assume that the nucleus is located at the position of the cross in Figure \ref{fig:slits}. The spatial positions along the three slit positions were registered relative to the nucleus by first fitting Gaussians to the peak emission in spatial continuum slices taken through the long-slit spectra. The offsets of these continuum centroids from the nucleus
were then measured using the line-free HST continuum image shown in Figure \ref{fig:slits}. In this way, it was possible to estimate the distances from the AGN of different positions along the slits.

\section{Results}

\subsection{Emission-line kinematics and spatial extent}

\begin{figure}
\includegraphics[width=9.7cm]{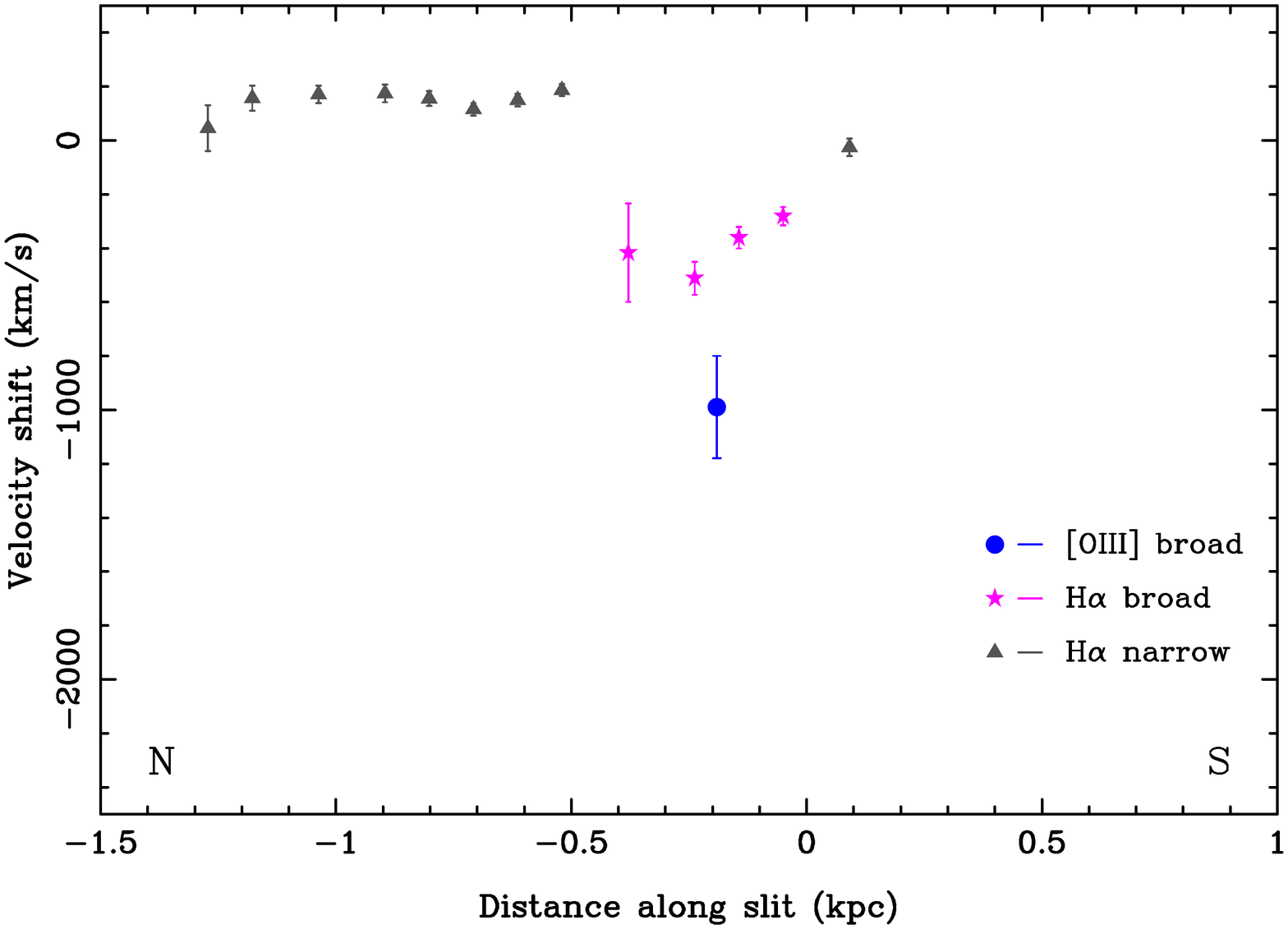}
\includegraphics[width=9.7cm]{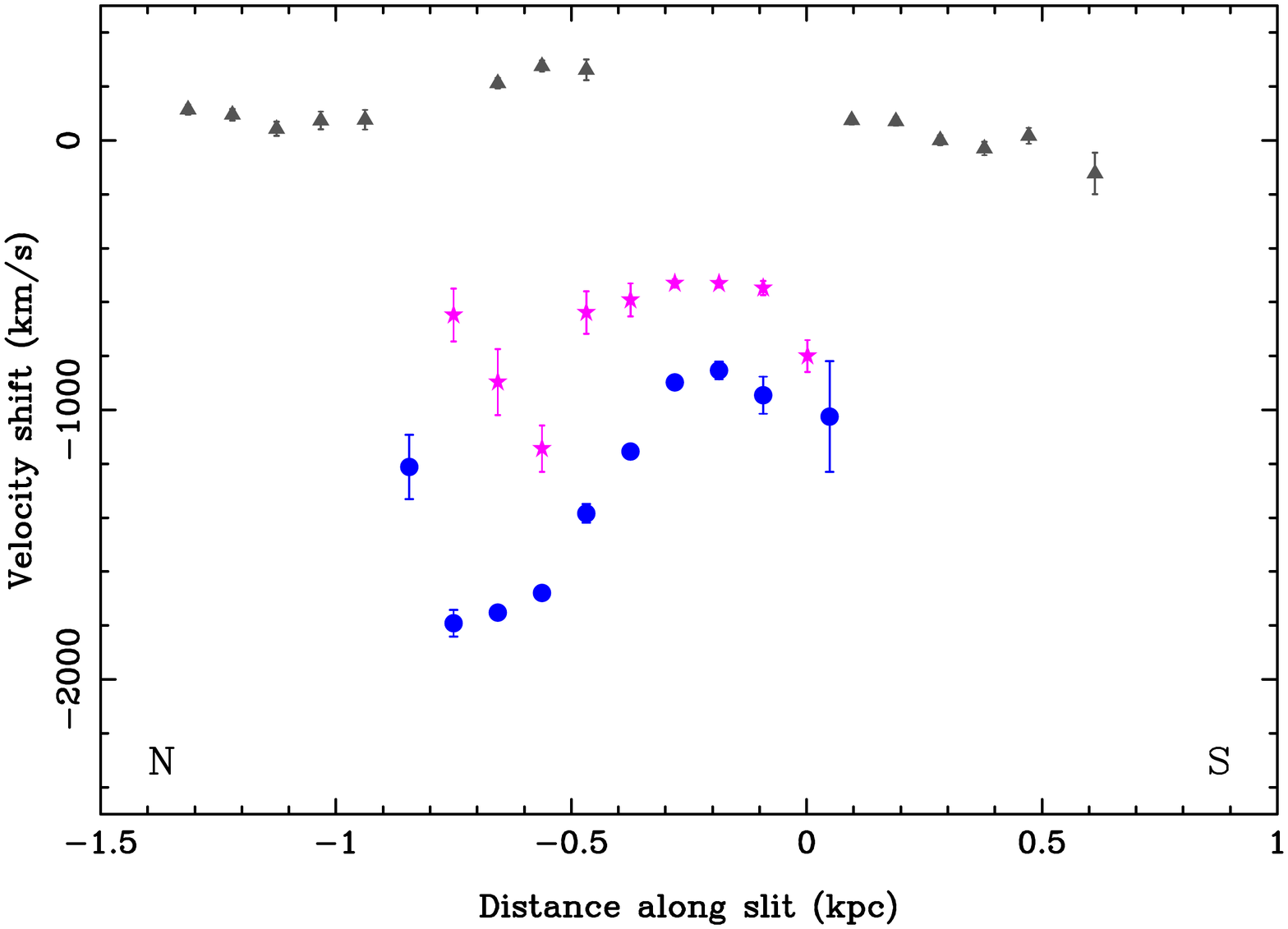}
\includegraphics[width=9.7cm]{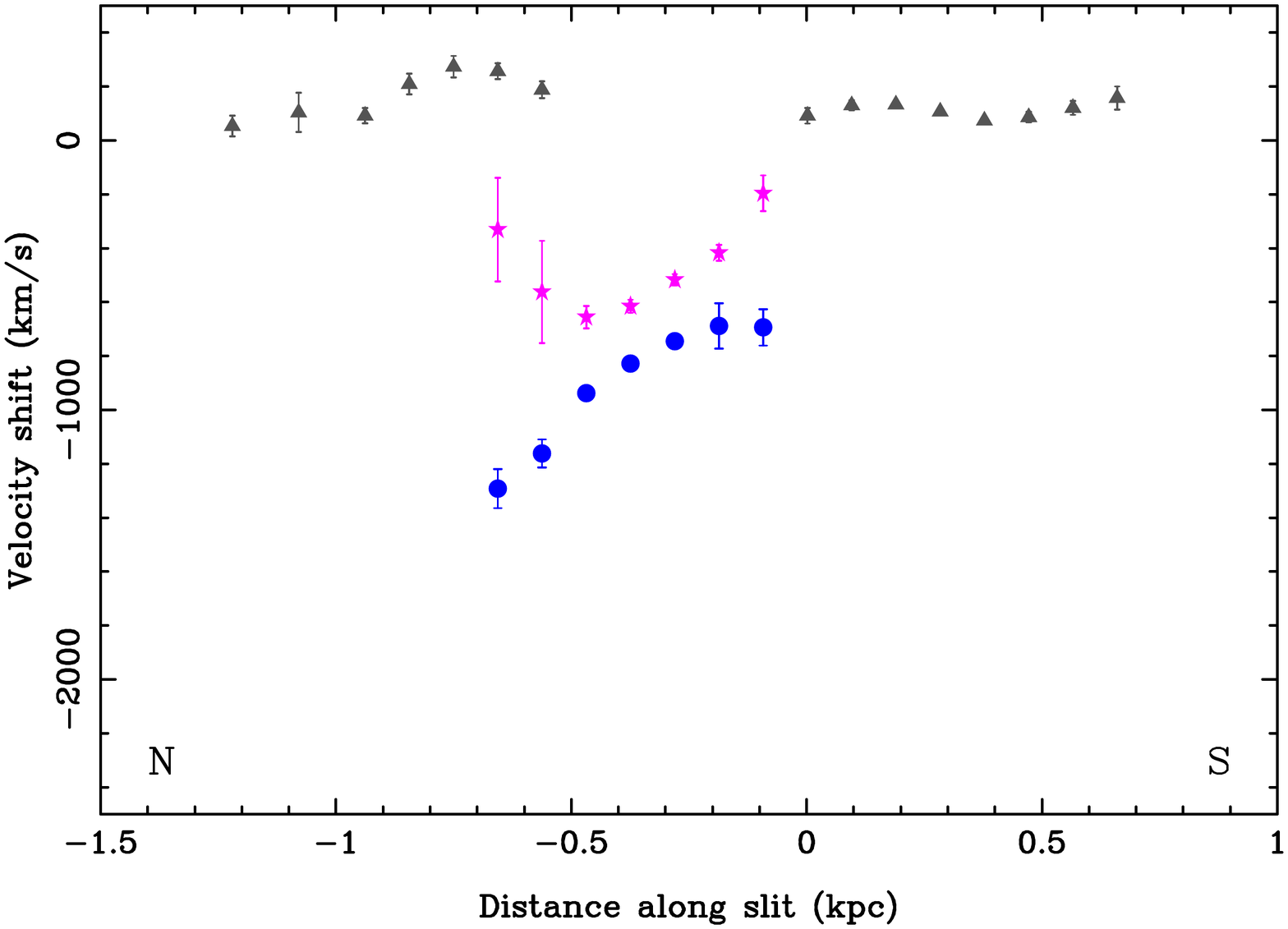}
\caption{The variation in velocity shifts as a function of spatial position along the east (top), centre (middle), and west (bottom) slits. In each case, the zero point of the x-axis represents the intercept of the slit and the line
drawn  horizontally from the assumed position of the AGN in Figure \ref{fig:slits}. The blue circles, pink stars, and grey triangles represent the measured
[OIII], H$\alpha$+[NII] broad, and H$\alpha$+[NII] narrow kinematics respectively. The velocity shifts are referred to the rest-frame of the host galaxy; north is to the left and south to the right.}
\label{fig:velocities}
\end{figure}

\begin{figure}
\includegraphics[width=9.7cm]{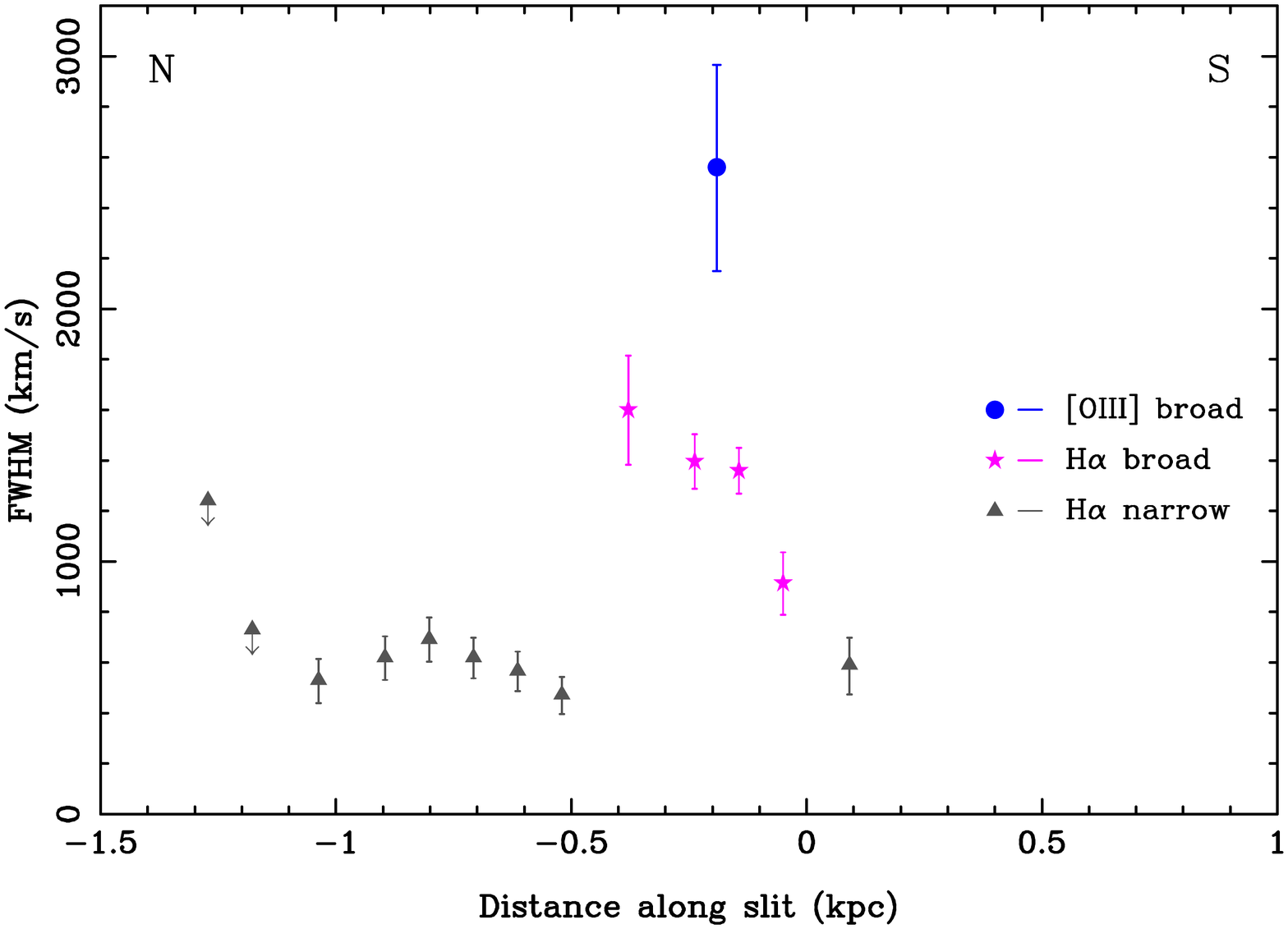}
\includegraphics[width=9.7cm]{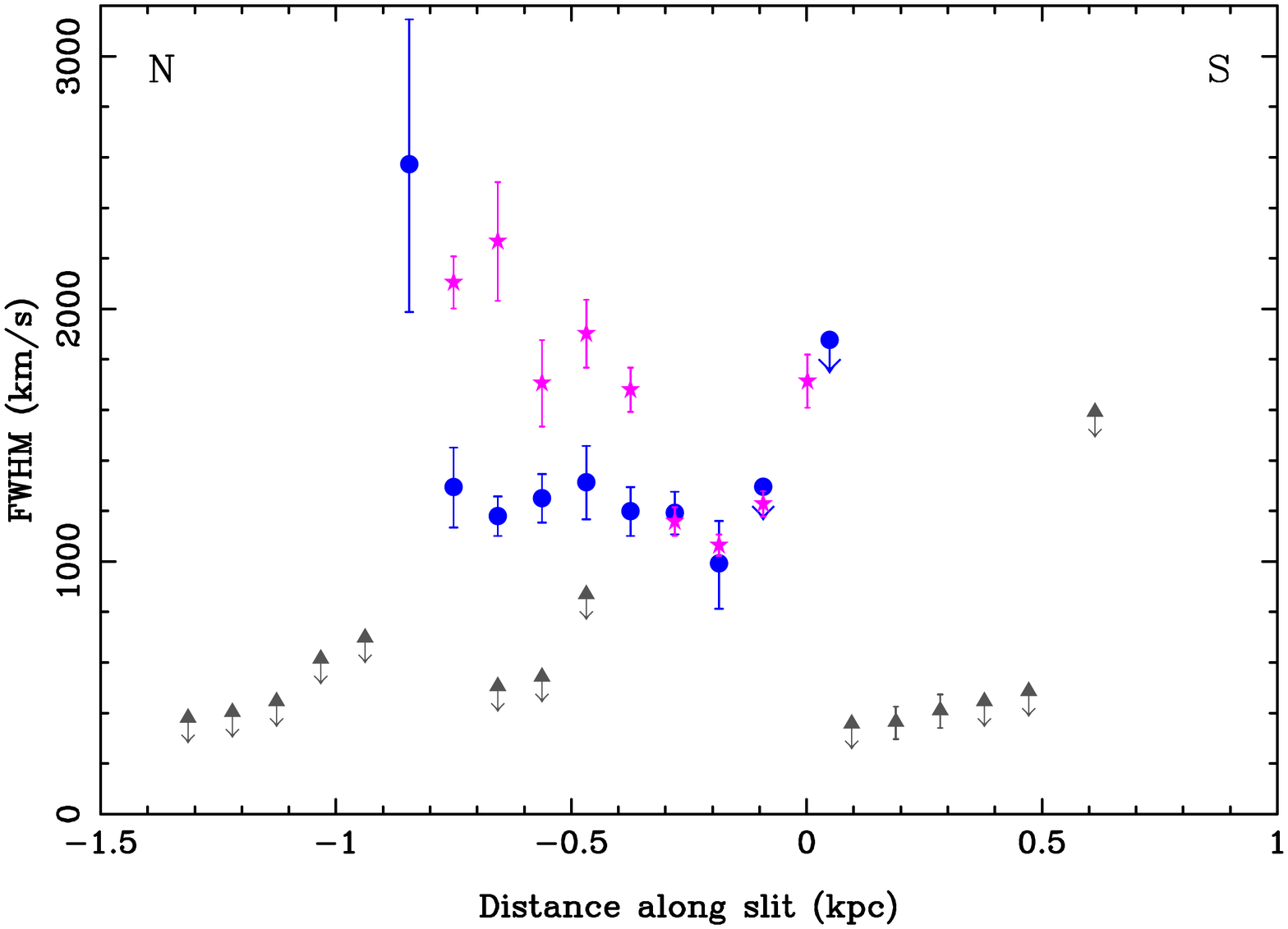}
\includegraphics[width=9.7cm]{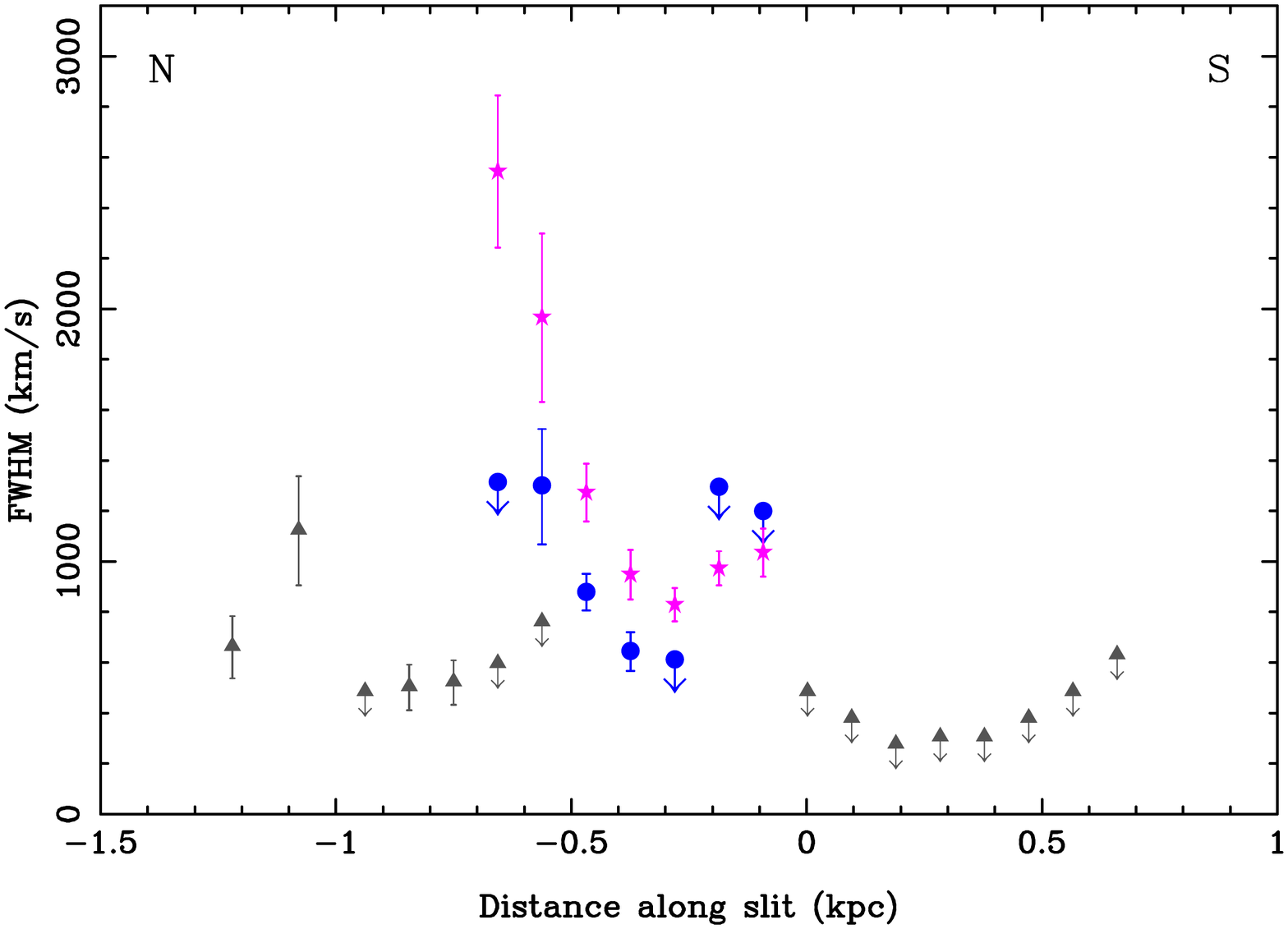}
\caption{The variation in velocity widths ($FWHM$) as a function of spatial position along the east (top). centre (middle), and west (bottom) slits. In each case, the zero point of the x-axis represents the intercept of the slit and the line
drawn horizontally from the assumed position of the AGN in Figure \ref{fig:slits}. The blue circles, pink stars and grey triangles represent the measured
[OIII]. H$\alpha$+[NII] broad and H$\alpha$+[NII] narrow kinematics respectively. All velocity widths have been corrected for
the instrumental width; north is to the left and south to the right.}
\label{fig:widths}
\end{figure}

Figure \ref{fig:2d} shows grey-scale images of the long-slit spectra in the region of H$\alpha$+[NII], and Figures \ref{fig:velocities} and \ref{fig:widths} present the radial velocity shifts ($\Delta V$) and line widths ($FWHM$) measured along all three slit positions. The velocity shifts are measured relative to the host galaxy redshift determined by \citet{spence18} using stellar absorption features detected in the nuclear spectrum of F14394+5332E, and the line widths have been corrected in quadrature for the assumed instrumental width (14.41$\pm$0.28 \AA); all velocity shifts and line widths have been corrected to the host galaxy rest frame.

Without binning the data in the spatial direction, the [OIII]$\lambda\lambda$5007,4959 lines are only detected in individual rows of the reduced 2D spectrum out to a radius of 0.47 arcsec (0.88\,kpc) to the north of the assumed AGN position, corresponding to the high surface brightness structure visible in our HST/ACS [OIII] narrow-band image. This compares with the flux-weighted mean radial extent of $0.65 < r < 0.84$\,kpc measured by \citet{tadhunter18} using HST [OIII] narrow-band imaging, and the $r=0.75\pm0.12$\,kpc measured by \citet{spence18} from the seeing-corrected spatial 
extent of the broad wings of the [OIII] emission in their ground-based long-slit spectrum. Note that the [OIII] lines are not clearly detected to the south of the AGN in individual rows of the STIS spectra\footnote{N.B. Although the [OIII] narrow-band HST image shows a more extensive nebulosity (maximum extent 2.1\,kpc: \citealt{tadhunter18}), the more extended [OIII] emission has too low surface brightness to be detected in individual rows of the HST/STIS spectra, given the exposure times used for the observations.}.

 In contrast, the H$\alpha$+[NII]
lines are detected in individual rows out to a significantly larger radius: 0.72 arcsec (1.34\,kpc) to the north and 
0.36 arcsec (0.66\,kpc) to the south of the AGN. Moreover, whereas the H$\alpha$+[NII] lines are well detected across several rows in the E slit, the [OIII] is barely detected in that slit; the [OIII] is also less extended in the W slit than it is in the C slit. 

We note several interesting features of the emission-line kinematics as follows.
\begin{itemize}
\item {\bf [OIII] kinematics.} Along the C slit the lines are broad ($FWHM > 950$ km s$^{-1}$) and blueshifted ($\Delta V < -700$ km s$^{-1}$) at all positions in which the [OIII] is detected. There is also a steep velocity gradient, with the magnitude of the blueshifted velocity increasing from $\sim$750 to 1800 km s$^{-1}$ from the inner to the outer parts of the outflow. A similar velocity gradient, but with lower amplitude, is observed along the W slit, and there is also evidence for lower line widths along that slit ($450 < FWHM < 1250$ km s$^{-1}$). The [OIII] is barely detected
along the E slit, apart from a compact region to the NE of the nucleus, where the [OIII] lines appear broad ($FWHM > 2000$ km s$^{-1}$), although the uncertainty is large in this region due to a relatively low S/N.
\item {\bf H$\alpha$+[NII] kinematics.} There is evidence for broad ($FWHM > 800$ km s$^{-1}$), blueshifted ($\Delta V < -200$ km s$^{-1}$) components to the H$\alpha$+[NII] lines in all the rows of the spectra in which the [OIII] is detected. Outside these regions, the H$\alpha$+[NII] lines are relatively narrow ($FWHM < 600$ km s$^{-1}$) -- often unresolved -- and show a small redshift relative to the host galaxy rest frame ($\Delta V < 300$ km s$^{-1}$). The region in which the narrow H$\alpha$+[NII] lines are detected overlaps with the more distant kinematically disturbed region to the north of the nucleus, where two Gaussian components (broad and narrow) are required to fit the lines.
\item {\bf [OIII] and H$\alpha$+[NII] comparison.} In the regions in which the [OIII] is detected in individual rows, the kinematics of the broader H$\alpha$+[NII] components follow those of the the [OIII] lines in the sense that
the lines are broad. However, the kinematics of the broader H$\alpha$+[NII] lines appear to differ in this region in the sense that the measured amplitude of the blueshifted velocities is lower, there is no evidence for the steep velocity gradient with radius observed in the [OIII] lines, and, at greatest radial distances from the nucleus, the lines appear broader. It is notable that the largest differences between the [OIII] and H$\alpha$+[NII] (broader component) kinematics are found in the region in which two kinematic components (broad and narrow) are required to fit the H$\alpha$+[NII] lines. This suggests that some of the differences between the [OIII] and H$\alpha$+[NII] may not be real, but rather related to degeneracies in the fit to the H$\alpha$+[NII] blend, given the low spectra resolution of the data.
\end{itemize}

Overall, the results are consistent with the those of our previous ground-based spectroscopy and HST imaging studies of F14394+5332E, in the sense that they confirm the presence of a relatively compact outflow region to the north of the nucleus. This outflow is set in a more extensive and kinematically quiescent halo that is only detected in H$\alpha$+[NII] in individual rows of the STIS spectra. The presence of this kinematically quiescent halo reinforces the view
that the AGN-driven outflow in this object has yet to encompass the full volume of warm/cool gas in the host galaxy.

Interestingly, in the kinematically disturbed regions closest to the nucleus ($r < 0.5$\,kpc) there is a ``hole'' in the distribution of the H$\alpha$+[NII] narrow component in all three slit positions. This suggests that, at this location, {\it all} the warm gas has been accelerated to high velocities. The fact that there is an overlap between the narrow and broad H$\alpha$+[NII] components at larger radial distances to the N of the AGN may then be explained as a line-of-sight projection effect i.e. the narrow components detected in this overlap region are emitted by warm gas in the foreground that is at larger true radial distances from the AGN than the outflow.

\subsection{Ionization}

\begin{figure}
\includegraphics[width=9.7cm]{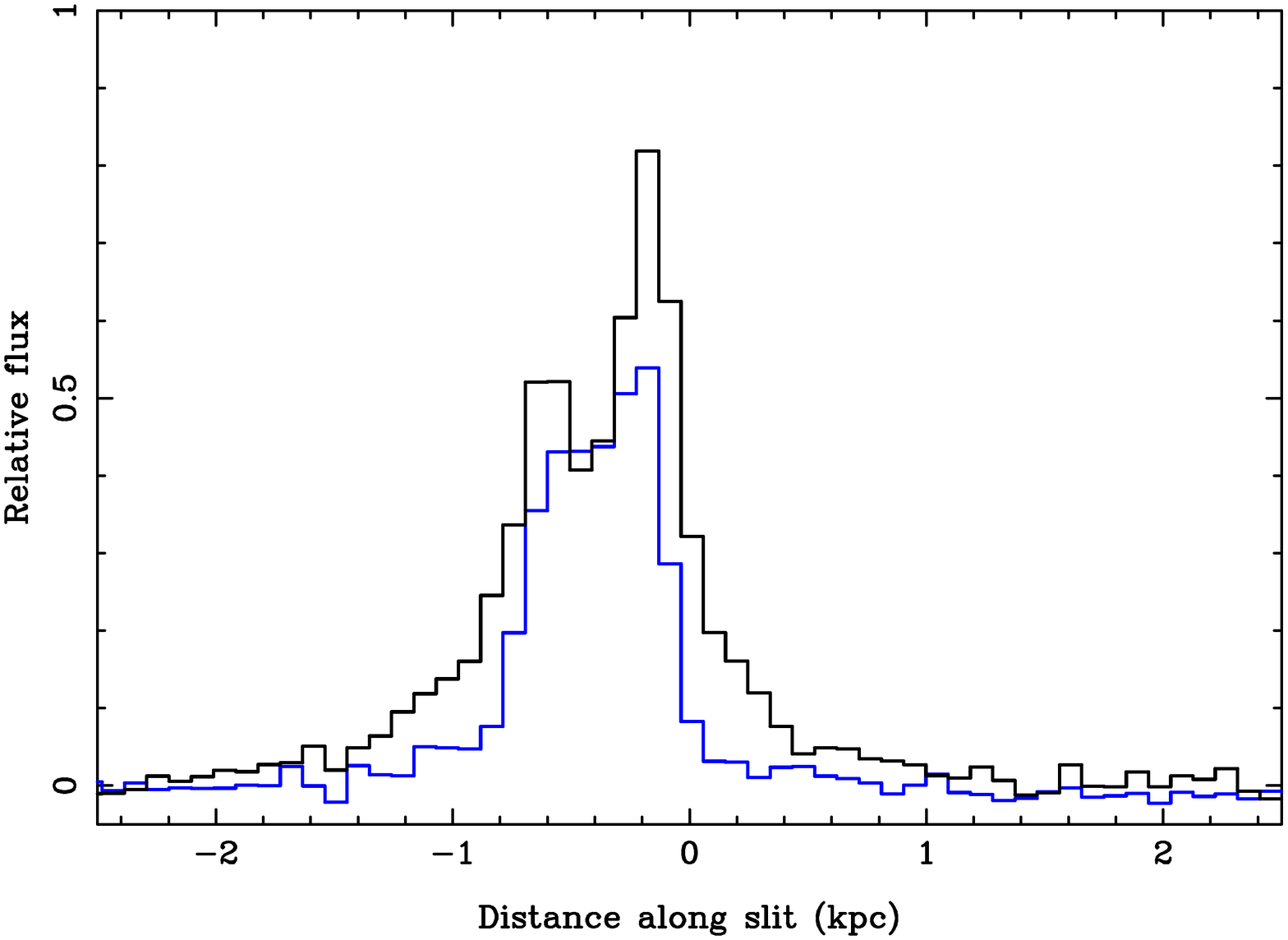}
\caption{The spatial distributions of [OIII]$\lambda\lambda$5007,4959 (blue) and H$\alpha$+[NII]$\lambda\lambda$6584,6548 (black), as measured using continuum-subtracted spatial slices extracted from the central long-slit spectrum. The spatial slice for the redshifted [OIII] lines was integrated over the
wavelength range 5425 -- 5557\,\AA, whereas that for the redshifted H$\alpha$+[NII] lines was integrated over the wavelength range 7176 -- 7321\,\AA; the wavelength range 5763 -- 6780\,\AA\, was used for the continuum slice that was scaled to the same wavelength range as the emission-line slices prior to subtraction.
Note the different spatial distributions of the 
[OIII] and H$\alpha$+[NII] emission.}
\label{fig:slices}
\end{figure}

Figure \ref{fig:slices} compares the spatial distributions of the integrated [OIII]$\lambda\lambda$5007,4959 and H$\alpha$+[NII]$\lambda\lambda$6584,6548 flux along the C slit. These were derived by extracting spatial slices from our long-slit spectra and subtracting a continuum slice extracted from an adjacent wavelength region of line-free continuum. The differences between the spatial distributions of the two emission-line blends are striking: the H$\alpha$+[NII] slice shows a double-peaked structure in the integrated emission-line flux that is associated with the region of kinematic disturbance (see also Figure \ref{fig:2d}), whereas this double-peaked structure is much less pronounced in the [OIII] slice; in the more extended, lower-surface-brightness regions outside the high-surface-brightness region that encompasses the  peaks  of flux distributions, the H$\alpha$+[NII] emission  is clearly much stronger than the [OIII] emission. This suggests that the degree of ionization of the warm gas peaks in the outflow, but is much lower in the extended halo. 

\begin{center}
\begin{figure}
\includegraphics[width=7.0cm]{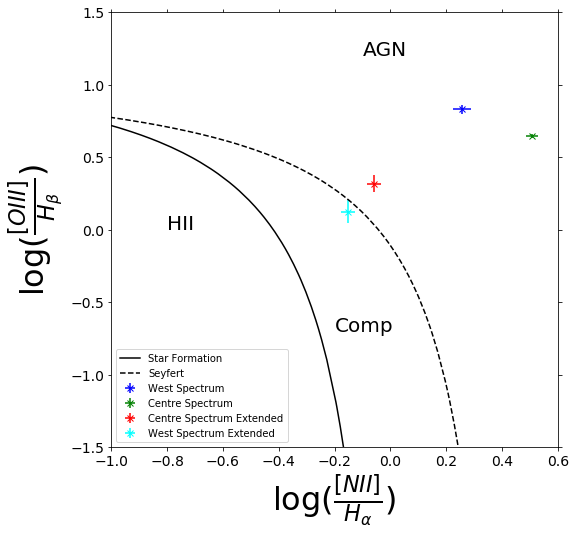}
\includegraphics[width=7.0cm]{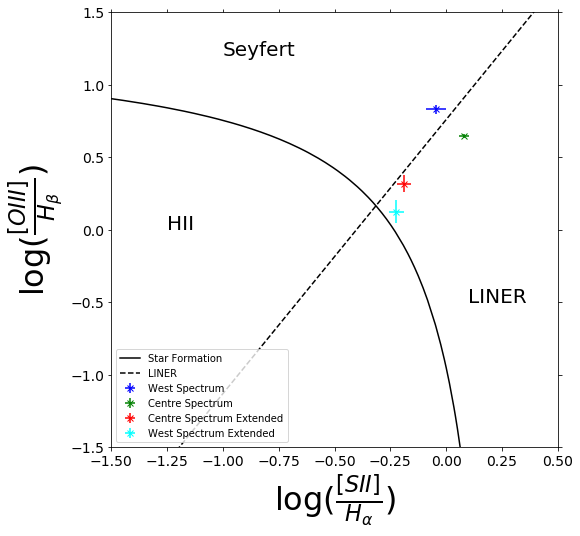}
\includegraphics[width=7.0cm]{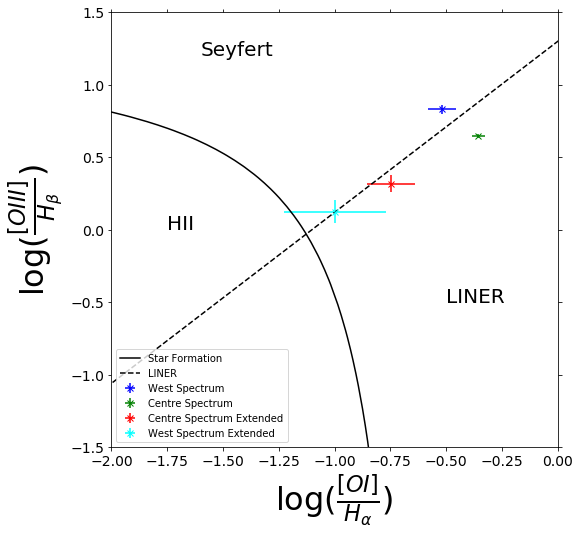}
\caption{Standard BPT diagnostic diagrams showing the measured line ratios for integrated spectra of the kinematically
disturbed central regions (dark-blue and green crosses) and kinematically quiescent extended regions (red and light-blue crosses), as measured from the central and western long-slit spectra. 
The spectra for the kinematically disturbed regions were extracted over the regions in which the [OIII] was clearly detected in individual rows of the long-slit spectra (corresponding to apertures of 0.1x0.51 arcsec and  0.1x0.36 arcsec for the central and western slits respectively), whereas the spectra of the kinematically undisturbed regions were extracted for regions further to the north and south along the slits in which the H$\alpha$+[NII] was detected in individual rows.
The lines that de-limit the Seyfert, HII and LINER zones of the
diagram and taken from \citep{kewley06}. 
}
\label{fig:bpt}
\end{figure}
\end{center}

To further investigate the ionization of the warm gas in the kinematically disturbed and quiescent regions along the W and C slits, we plot the emission line ratios measured from the spatially-integrated STIS spectra of these regions on the standard Baldwin, Philips \&
Terlevich (BPT) diagrams \citep{bpt,kewley06} shown in Figure \ref{fig:bpt}. The trend is consistent across all three of the diagrams: while the line ratios for all the spatial regions fall in the AGN/LINER parts of the diagrams, 
the ratios for the kinematically quiescent regions fall closer to the HII/composite zone than those for the kinematically disturbed regions. This trend can be explained in two ways: (a) the kinematically-disturbed regions are dominated by AGN ionization (AGN photoionisation or shocks), whereas the line ratios of the  kinematically-quiescent regions reflect a mixture of AGN and stellar ionization \citep[e.g.][]{santoro16}; or (b) both the kinematically disturbed and undisturbed regions are photoionized by the AGN, but the gas abundances are higher and/or the AGN ionising radiation field harder in the kinematically disturbed regions \citep[see][]{rz13,santoro18}.  However, it is not possible to determine the dominant ionization mechanism for the warm gas from these diagrams alone, because the measured ratios fall in regions of the BPT diagrams in which the shock and AGN photoionization model grids show a high degree of overlap. 

\section{Discussion}

By directly confirming the compactness of the outflow region in F14394+5332E, the results presented in this paper further reinforce the idea that the warm, AGN-driven outflows in nearby ULIRGs are generally compact \citep{rose18,tadhunter18,spence18}, despite these being some of the most actively evolving galaxies in the local Universe; they are also consistent with recent results based on higher spectral resolution STIS spectroscopy observations of nearby quasar 2 sources \citep{fischer18,revalaski18}. We now discuss the implications of the results for understanding the gas acceleration mechanism, and for calculations of the mass outflow rates and kinetic powers of the outflows.

\subsection{The origin of the emission-line kinematics}

In terms of the acceleration mechanism for the warm gas, it is interesting that the emission lines are not only highly blueshifted but broad throughout the outflow region, and that the [OIII] emission lines maintain a uniform line width along the C slit, where the lines are best detected. We note that these kinematics are unlikely to arise through gravitational motions in the merger: not only is the detection of such extreme near-nuclear emission line kinematics in ULIRGs almost invariably associated with the subset of ULIRGs hosting optical AGN nuclei \citep{rz13,arribas14} -- providing strong evidence that the outflows are AGN-driven -- but integral field observations of the warm ionised gas in large samples of nearby luminous- and ultra-luminous infrared merger systems show no evidence for such high velocity shifts and large line widths on kpc scales \citep[e.g.][]{bellocchi13}. 
Moreover, hydrodynamic simulations of major galaxy mergers fail to produce gas velocities of more than a few 100\,km s$^{-1}$ unless AGN outflows are included \citep[e.g.][]{debuhr12}.

Although the steep increase in the (projected)  blueshift of the [OIII] emission line with radial distance from the nucleus along the C and W slits (Figure \ref{fig:velocities}) might at first sight be taken as evidence for a radially accelerating outflow, it seems unlikely that individual clouds would survive being accelerated from the nucleus to their current position without being destroyed by the hydrodynamic instabilities associated with the acceleration process. This is especially the case if the warm gas has a low filling factor and is being swept up in the high filling factor hot gas behind a rapidly expanding outer shock, itself driven by the inner (faster, hotter) AGN wind \citep{king15}. 

More plausibly, the clouds are being accelerated {\it locally} by the outer shock, and the large line widths -- suggestive of a highly turbulent medium -- are signalling the clouds' hydrodynamic interaction via Kelvin-Helmholtz instabilities with the hot, post-shock gas \citep[e.g.][]{klein94,mellema03}. The apparent velocity gradient could then reflect the fact that the outer clouds are easier to accelerate (e.g. due to having lower densities) by this mechanism than the inner clouds. 

Alternatively, the velocity gradient might reflect a systematic change in the direction of the outflow velocity vector relative to the line of sight (LOS) with radius.  For example, if we imagine that the outflowing gas forms part of a cone of outflowing material to the  N of the AGN, with the warm gas having a clumpy distribution within the cone, then the regions at small radial distance from the AGN could fall in part of the cone that is closer to the plane of the sky (velocity vector directed further from the LOS), and the regions at larger radial distance from the AGN in the part of the cone that is closest to the observer (velocity vector closer to the LOS). In this case, the [OIII] velocities measured at the largest radial distances from the AGN would be more representative of the true (de-projected) velocity of the warm outflow.

Finally, we should consider the possibility that the warm clouds somehow avoid destruction via interaction with the hot wind, and the velocity gradient reflects a true accelerating outflow (clouds accelerated from the nucleus to their current position). In this case, the large line widths might be due to the range of projections of the velocity vector for components at a wide range of different depths in the outflow cone along the LOS, rather than a high local velocity dispersion of a relatively small volume within the cone. However, this possibility seems unlikely for two reasons. First, in the (likely) event that the warm gas in the outflow cone has a clumpy, non-uniform distribution, it would be difficult to explain the lack of variation in the [OIII] line width along  C slit position (Figure \ref{fig:2d}) if the large line widths were mainly due to the range of velocity vector projections along any particular LOS. Second, if the gas were accelerating with distance from the nucleus and the line widths were entirely due to projection effects, we would expect not only the projected blueshifted velocity, but also the line width, to increase with radius, which is not observed for the [OIII] emission line\footnote{Although the H$\alpha$ and [NII] lines appear to show an increase in line width with radial distance from the AGN along the C slit, as argued previously, this may not truly reflect the emission line kinematics, but rather degeneracies in the fitting of multiple Gaussian components to the H$\alpha$+[NII] blend at low spectral resolution.}.

Therefore, it seems most plausible that the warm gas in the outflow has a high local velocity dispersion throughout the outflow region due to hydrodynamical interaction with the hot gas behind the outer shock as the clouds are accelerated, and that the apparent velocity gradient either reflects a systematic change in the direction of the velocity vector relative to the line of sight, or in the ease of acceleration of the clouds with distance from the nucleus. The corollary of this is that the sharply defined outflow region detected in our data represents the full extent of the outer shock driven by the AGN; if the shock extended further, then we would expect this to be reflected in the kinematics of the H$\alpha$+[NII] at larger radii, which is not the case. The only way to avoid this conclusion would  be if this quiescent gas were in reality at much larger radial distances, and only appeared within a few kpc of the nucleus due to projection effects, or if the outflow were highly collimated and the extended warm and cool gas were distributed away from the axis of the outflow.

\subsection{Determining masses and energetics}

Many previous attempts to quantify the warm outflows in nearby ULIRGs and similar objects were based on spatially-integrated
properties, and used single estimates of the outflow radius, density and emission-line luminosity. 
  
A particular problem
is that we measure projected, rather than true outflow velocities. In papers I and III we attempted to account for such 
projection effects by making two extreme assumptions about the projection factors: (i) we assumed that the mean velocity shifts of the emission-line features relative to the rest frame represents the true outflow velocity, and the line widths represent local velocity dispersion; alternatively (ii) we assumed that the velocity in the blue wing at which the profile contains 5\% of the total emission-line flux (v$_{05}$) represents 
the true outflow velocity, and the line widths are entirely due to the range of projections of the velocity vector within the unresolved outflow region. As discussed in paper I,  assumption (i) is likely to lead to the mass outflow rate ($\dot{M}$) and kinetic power ($\dot{E}$) of the outflow being underestimated, since projection effects have not been taken into account, whereas assumption (ii) may lead to lead to overestimates if there is a substantial level of local velocity dispersion in the gas motions. Since our STIS observations suggest that, in fact, the warm gas does show a large line width  ($FWHM >$600\,km s$^{-1}$) throughout the outflow region, they are most consistent with the explanation for the line widths under assumption (i). 

For an extended outflow, the mass outflow rate for a particular resolved region of the outflow of measured (projected) radial extent $\Delta r$ can be estimated using:
\begin{equation}
\dot{M} = \frac{m_p L(H\beta) v_{out}}{\alpha^{eff}_{H\beta} h \nu_{H\beta} n_e \Delta r}
\end{equation}
where $L(H\beta)$ is the extinction-corrected H$\beta$ luminosity, $v_{out}$
is the measured (projected) outflow velocity, $n_e$ is the electron density, $m_p$ is the proton mass, $\alpha^{eff}_{H\beta}$ is the effective Case B H$\beta$ recombination coefficient\footnote{Here we assume $\alpha^{eff}_{H\beta} = 3.03\times10^{-14}$ cm$^3$ s$^{-1}$, which is appropriate for Case B recombination and an electron temperature of $T = 10^4$\,K and our assumed electron density \citep{osterbrock06}.}, $h$ is the Planck constant, and $\nu_{H\beta}$  the frequency of an H$\beta$ photon. Note that this equation does not take into account the effects of projection on the outflow velocity ($v_{out}$) nor on the radial extent of the extended outflow region ($\Delta r$). 

To evaluate the mass outflow rate at a given location we estimated $L(H\beta)$ using the measured H$\alpha$ fluxes, since H$\beta$ is too weak to be measured accurately in many individual rows of the long-slit spectrum. We first corrected the H$\alpha$ fluxes for extinction using the spatially averaged reddening value ($E(B-V)=0.63$ mag) estimated for the near-nuclear regions of F14394+5332E by \citet{spence18} using the trans-auroral line ratios, then converted to the H$\beta$ flux using the Case B recombination ratio H$\alpha$/H$\beta=2.85$, before calculating $L(H\beta)$ using the luminosity distance appropriate for our assumed cosmology. As well as a uniform extinction, we also assumed a uniform  electron density of $n_e=3500$\,cm$^{-3}$ in all spatial regions. The latter estimate is based on the trans-auroral emission line ratios for the broad emission lines measured in the nuclear regions by \citet{spence18}. Finally, to estimate of $v_{out}$ we used the velocity shifts measured relative to the galaxy rest frame for the broad H$\alpha$ components from the multi-Gaussian fits to the H$\alpha$+[NII] blend (pink stars in
Figure \ref{fig:velocities}). 

\begin{figure}
\includegraphics[width=10.0cm]{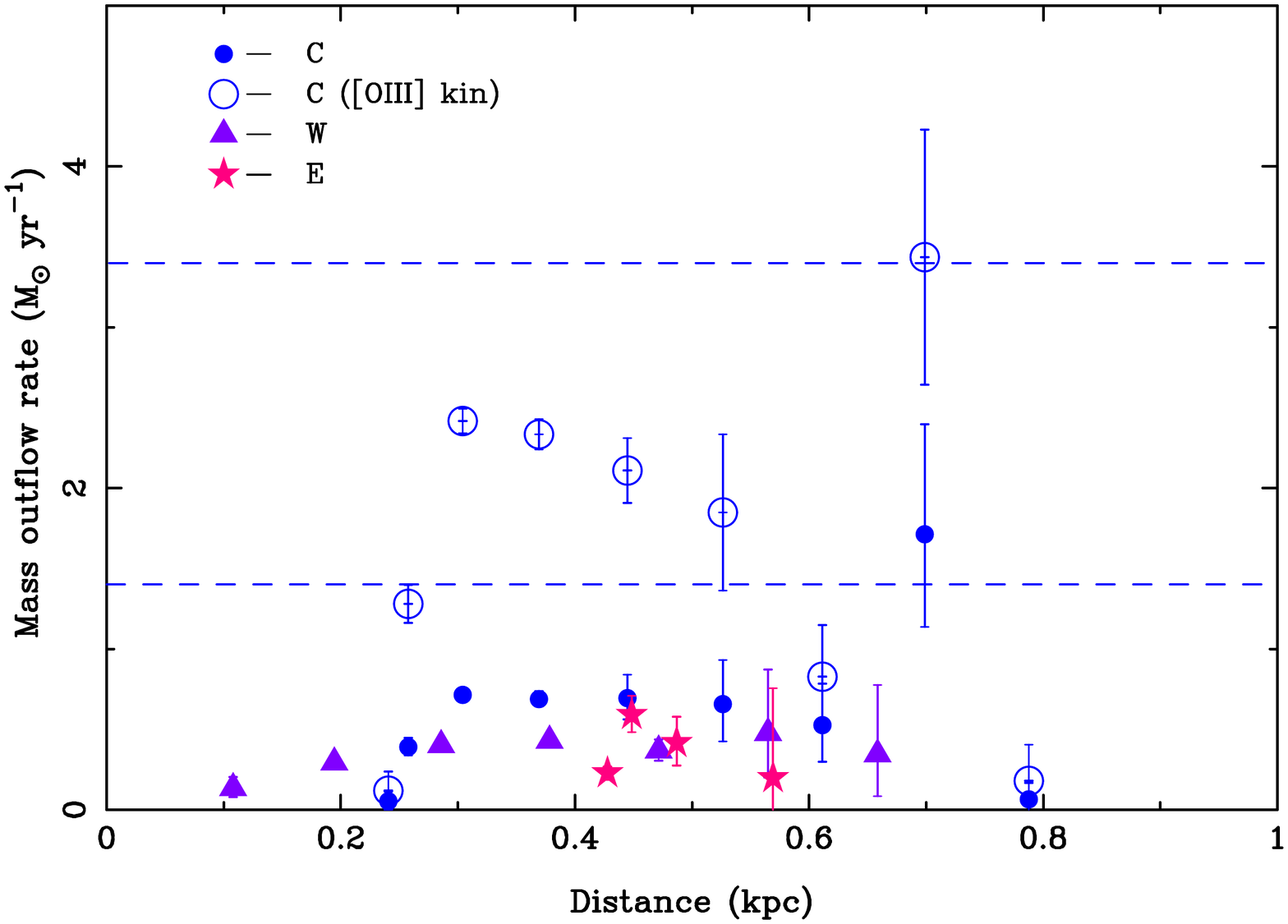}
\caption{Mass outflow rate as a function of radial distance from the nucleus, as estimated from the H$\alpha$ data
for different spatial locations along the eastern (pink stars), central (solid blue circles) and western (purple triangles) long-slit spectra. The open blue circles represent estimates made for the central slit position by using  maximal [OIII] kinematics with H$\alpha$ fluxes. The blue dashed lines show the spatially-integrated results from \citet{spence18} calculated
under assumptions (i) (bottom line) and (ii) (top line).
See the text for details.}
\label{fig:mdot}
\end{figure}

\begin{figure}
\includegraphics[width=10.0cm]{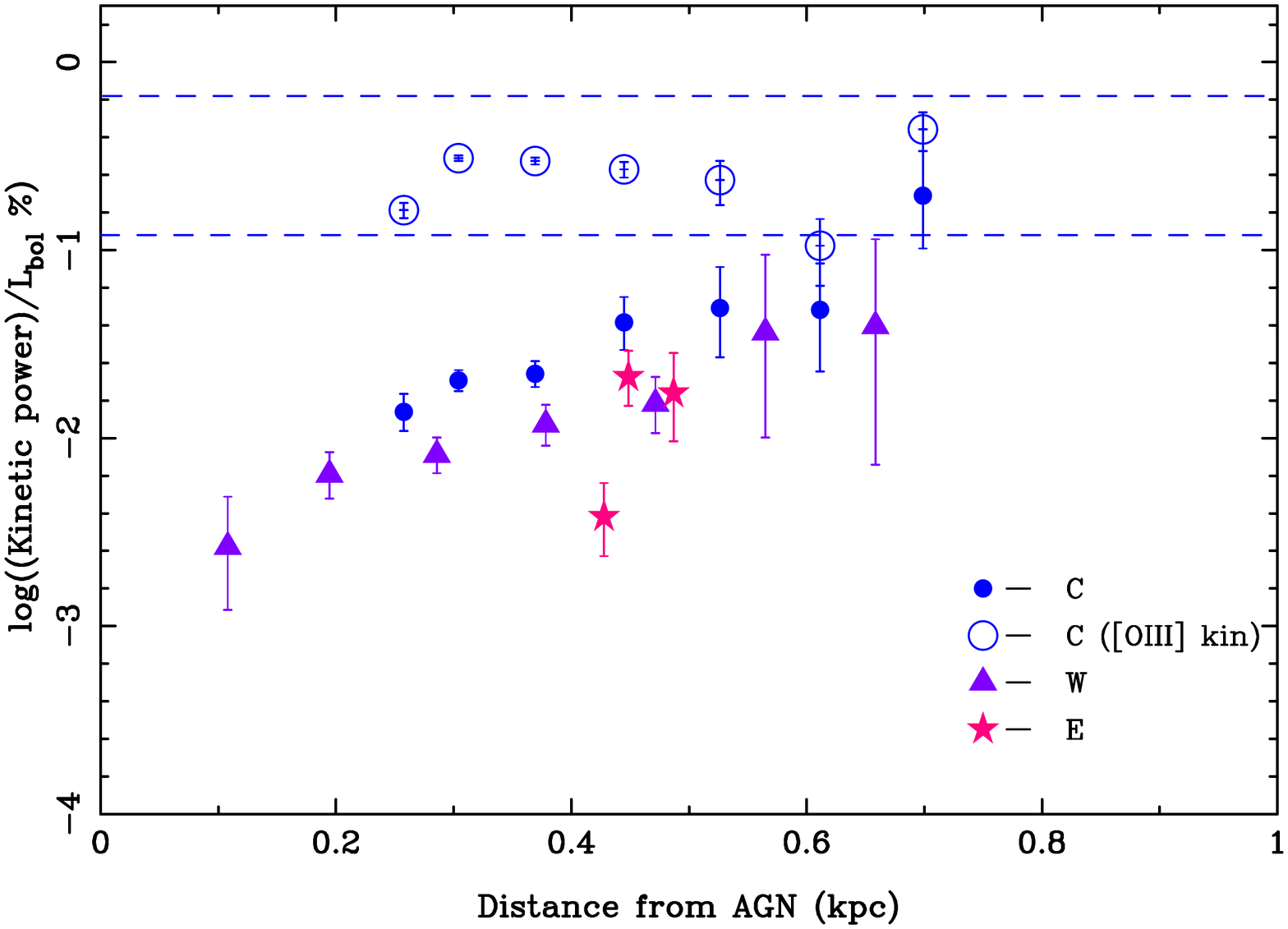}
\caption{The ratio of kinetic power and AGN bolometric luminosity as function of radial distance from the nucleus. The estimates are based on the H$\alpha$ data
for different spatial locations along the eastern (pink stars), central (solid blue circles) and western (purple triangles) long-slit spectra. The open blue circles represent estimates made for the central slit position by using  maximal [OIII] kinematics with H$\alpha$ fluxes. The blue dashed lines show the spatially-integrated results from \citet{spence18} calculated
under assumptions (i) (bottom line) and (ii) (top line). See the text for details.}
\label{fig:edot}
\end{figure}

The resulting mass outflow rates measured for all three slit positions are shown in Figure \ref{fig:mdot} (filled symbols). In the regions where the mass outflow rate is best measured, the typical values fall in the range $0.3 < \dot{M} < 0.7$\,M$_{\odot}$ yr$^{-1}$, with evidence that the mass outflow rates are systematically higher in the C, than in the E or W slits. For comparison, using spatially integrated emission-line measurements and the same density and reddening estimates \citet{spence18} obtained $\dot{M} = 1.4$\,M$_{\odot}$ yr$^{-1}$ assuming method (i) of that paper. Two factors could contribute the fact that that the HST/STIS  estimates are approximately a factor $\times2$  lower: first, in a given spatially resolved region along the spectroscopic slit we do not sample
all the gas at that radial distance from the nucleus; second, the H$\alpha$ velocity shifts for the broad H$\alpha$ lines are generally lower than those measured for the [OIII] lines at the same spatial locations and the $\Delta V = -990$\,km s$^{-1}$ assumed by \citet{spence18}. These factors, along with the fact that we have not corrected for projection effects mean that the HST/STIS $\dot{M}$ values calculated in this
way are likely to represent underestimates of the true values.

In order to calculate upper limits for $\dot{M}$, we avoid the effects of the potential degeneracies in the H$\alpha$+[NII] fits by assuming that the true outflow velocities of all emission lines are the same as those of [OIII]. Moreover, we attempt to account for projection effects by assuming that the true outflow velocity takes its maximum measured value in all spatial regions: $v_{out}=1800$\,km s$^{-1}$ --- the shift of [OIII] lines in the part of the outflow in the central slit that is most distant from the nucleus (see Figure \ref{fig:velocities}), which we argued in section 4.1 is most likely to reflect the true (deprojected) outflow velocity. The results for the central slit are shown by the open symbols in Figure \ref{fig:mdot}. These less conservative estimates are typically a factor 2 -- 3$\times$ higher than those estimated before:   $\dot{M} \sim$1 -- 3\,M$_{\odot}$ yr$^{-1}$ , straddling the spatially-integrated results obtained under assumption (i) ($\dot{M} = 1.4$\,M$_{\odot}$ yr$^{-1}$) and assumption (ii) ($\dot{M} = 3.4$\,M$_{\odot}$ yr$^{-1}$) by \citet{spence18}. Note, however, if we were to correct for the projection of the radial extent of the spatial aperture ($\Delta r$), the values would be lower than those shown in Figure \ref{fig:mdot}, since the true radial extents of the apertures must be larger than the observed ones. Therefore, the open symbols in Figure \ref{fig:mdot} are likely to represent upper limits.

Also, under assumption (i), the kinetic power in the outflow can be estimated using (see paper I):
\begin{equation}
\dot{E} = \frac{1}{2} \dot{M} (v^2_{out} + 0.54 FWHM^2)
\end{equation}
where $FWHM$ is the measured velocity full-width at half maximum of the emission line in the resolved region corrected for the instrumental width. To compare
the kinetic power estimates with the radiative power of the AGN, we divide the $\dot{E}$ values by the bolometric luminosity
of the AGN in F14394+5332E\footnote{Here we assume $L_{bol} = 10^{45}$\,erg s$^{-1}$, consistent with the [OIII] luminosity based estimates of \citep{spence18}.} to obtain the coupling efficiencies: $\dot{F}=\dot{E}/L_{bol}$. If we assume that $v_{out}$ and $FWHM$ are given by the 
measured shift and $FWHM$ of the broad H$\alpha$ component, the results are shown by the solid symbols in Figure \ref{fig:edot}, whereas less conservative values (open symbols) are obtained by assuming $v_{out}=1800$\,km s$^{-1}$ and $FWHM = 1200$\,km s$^{-1}$ -- consistent with the [OIII] kinematics in the central slit in the part of the outflow at the largest radial distance from the AGN. The measured coupling efficiencies are in the range $0.0025 < \dot{F} < 0.5$\%, with values obtained by assuming the maximal [OIII] kinematics falling close to the upper end of this range. For comparison, based on their spatially-integrated measurements, \citet{spence18} obtained  values $\dot{F} = 0.12$ and $\dot{F} = 0.66$\% under assumptions (i) and (ii) respectively.

Overall, our STIS results for F14394+5332E demonstrate that the failure to take into account geometrical effects in
the spatially-integrated estimates of $\dot{M}$, $\dot{E}$ and $\dot{F}$ is unlikely to lead 
to the true values being over- or under-estimated by
more than a factor $\sim$2 -- 3. 

\section{Conclusions}

Our STIS/HST results for F14394+5332E provide further evidence that the warm ionized outflows in ULIRGs are compact and do not encompass the full radial extent of warm and cool phases of the ISM in the host galaxies. The large line widths that we measure throughout the outflow region are consistent with the idea that the warm gas in the outflow has been accelerated by the forward shock driven by the AGN, and that we are witnessing the hydrodynamic interaction of dense clumps of gas  with the wind of
hot gas behind the shock. Moreover, even with the new constraints on the spatial distribution of the outflow, the
estimated mass outflow rates and kinetic powers relative to the bolometric luminosity of the AGN remain modest, in line with our previous work in papers I \& III.

There is increasing evidence that much of the mass and power in the AGN-driven outflows in ULIRGs and similar objects is carried by the neutral and molecular gas phases, although the acceleration mechanism for these cooler gas phase outflows is uncertain. If the molecular outflows form as the denser gas cools behind the forward shock \citep{tadhunter14,morganti15,richings18}, then we predict that any molecular gas outflow in F14394+5332E will have a radial extent that is similar to, or smaller than, the warm gas outflow. In addition, we expect to detect X-ray emission from the hot gas behind the forward shock that accelerates and shreds the dense clouds. Therefore, future mm-wave CO and X-ray observations of F14394+5332E and similar objects will be
important for understanding the acceleration of the multi-phase outflows driven by AGN as their black holes grow rapidly at the peaks of major galaxy mergers.

\

\section*{Acknowledgements}
MR \& CT acknowledge support from STFC, and CRA acknowledges the Ram\'{o}n y Cajal Program of the Spanish Ministry of Economy and Competitiveness through project RYC-2014-15779 and the Spanish Plan Nacional de Astronom\'{i}a y Astrofis\'{i}ca under grant AYA2016-76682-C3-2-P. Based on observations taken with the NASA/ESA Hubble Space Telescope, obtained at the Space Telescope Science Institute (STScI), which is operated by AURA, Inc. for NASA under contract NAS5-26555. The authors acknowledge the data analysis facilities provided by the Starlink Project, which was run by CCLRC on behalf of PPARC. This research has made use of the NASA/IPAC Extragalactic Database (NED) which is operated by the Jet Propulsion Laboratory, California Institute of Technology, under contract with the National Aeronautics and Space Administration. We thank Michelle Berg, Marco Chiaberge, Javier Rodr\'iguez Zaur\'in, Marvin Rose and Henrik Spoon for their contributions to the early stages of this project.












\bsp	
\label{lastpage}
\end{document}